\documentclass[5p]{elsarticle}

\usepackage{hyperref}
\usepackage{amsmath, amssymb, amsfonts, amsthm}

\journal{Nuclear Instruments and Methods in Physics Research Section A}

\begin{document}
\begin{frontmatter}

\title{Detection System for Neutron $\beta$ Decay Correlations in the UCNB and Nab experiments}

\author[lanl,ornl]{L.~J.~ Broussard\corref{mycorrespondingauthor}}
\cortext[mycorrespondingauthor]{Corresponding author}
\ead{broussardlj@ornl.gov}
\author[lanl,ncsu]{B.~A.~Zeck}
\author[indiana]{E.~R.~Adamek}
\author[uva]{S.~Bae$\ss$ler}
\author[utk]{N.~Birge}
\author[lanl,cleve]{M.~Blatnik}
\author[ornl]{J.~D.~Bowman}
\author[lanl,ncsu]{A.~E.~Brandt}
\author[uky]{M.~Brown}
\author[lanl]{J.~Burkhart}
\author[indiana]{N.~B.~Callahan}
\author[lanl]{S.~M.~Clayton}
\author[uky]{C.~Crawford}
\author[ncsu]{C.~Cude-Woods}
\author[lanl]{S.~Currie}
\author[ncsu]{E.~B.~Dees}
\author[vt]{X.~Ding}
\author[utk]{N.~Fomin}
\author[uva]{E.~Frlez}
\author[uva]{J.~Fry}
\author[reg]{F.~E.~Gray}
\author[uky]{S.~Hasan}
\author[caltech]{K.~P.~Hickerson}
\author[ncsu]{J.~Hoagland}
\author[tenntech]{A.~T.~Holley}
\author[lanl]{T.~M.~Ito}
\author[lanl]{A.~Klein}
\author[uva]{H.~Li}
\author[indiana]{C.-Y.~Liu}
\author[lanl]{M.~F.~Makela}
\author[lanl]{P.~L.~McGaughey}
\author[lanl]{J.~Mirabal-Martinez}
\author[lanl]{C.~L.~Morris}
\author[lanl,uw]{J.~D.~Ortiz}
\author[lanl]{R.~W.~Pattie,~Jr.}
\author[ornl]{S.~I.~Penttil{\"a}}
\author[uky]{B.~Plaster}
\author[uva]{D.~Po{\v c}ani{\' c}}
\author[lanl]{J.~C.~Ramsey}
\author[uva]{A.~Salas-Bacci}
\author[lanl,indiana]{D.~J.~Salvat}
\author[lanl]{A.~Saunders}
\author[lanl]{S.~J.~Seestrom}
\author[lanl]{S.~K.~L.~Sjue}
\author[uky]{A.~P.~Sprow}
\author[lanl]{Z.~Tang}
\author[vt]{R.~B.~Vogelaar}
\author[ncsu]{B.~ Vorndick}
\author[lanl]{Z.~Wang}
\author[lanl]{W.~Wei}
\author[ncsu]{J.~Wexler}
\author[lanl]{W.~S.~Wilburn}
\author[lanl]{T.~L.~Womack}
\author[ncsu]{A.~R.~Young}

\address[lanl]{Los Alamos National Laboratory, Los Alamos, NM  87545  USA}
\address[ornl]{Oak Ridge National Laboratory, Oak Ridge, TN 37831  USA}
\address[ncsu]{North Carolina State University, Raleigh, NC 27695  USA}
\address[indiana]{Indiana University, Bloomington, IN  47405  USA}
\address[uva]{University of Virginia, Charlottesville, VA 22904  USA}
\address[utk]{University of Tennessee, Knoxville, TN 37996  USA}
\address[cleve]{Cleveland State University, Cleveland, OH  44115  USA}
\address[uky]{University of Kentucky, Lexington, KY 40506  USA}
\address[vt]{Virginia Polytechnic Institute \& State University, Blacksburg, VA 24061  USA}
\address[reg]{Regis University, Denver, CO  80221  USA}
\address[caltech]{California Institute of Technology, Pasadena, CA 91125  USA}
\address[tenntech]{Tennessee Technological University, Cookeville, TN  38505  USA}
\address[uw]{University of Washington, Seattle, WA  98195  USA}

\begin{abstract}
We describe a detection system designed for precise measurements of angular correlations in neutron $\beta$ decay.  The system is based on thick, large area, highly segmented silicon detectors developed in collaboration with Micron Semiconductor, Ltd.  The prototype system meets specifications for $\beta$ electron detection with energy thresholds below 10~keV, energy resolution of $\sim$3~keV FWHM, and rise time of $\sim$50~ns with 19 of the 127 detector pixels instrumented. Using ultracold neutrons at the Los Alamos Neutron Science Center, we have demonstrated the coincident detection of $\beta$ particles and recoil protons from neutron $\beta$ decay.  The fully instrumented detection system will be implemented in the UCNB and Nab experiments, to determine the neutron $\beta$ decay parameters $B$, $a$, and $b$.

\end{abstract}


%\footnotetext[1]{NOTICE OF COPYRIGHT \\This manuscript has been authored by UT-Battelle, LLC under Contract No. DE-AC05-00OR22725 with the U.S. Department of Energy. The United States Government retains and the publisher, by accepting the article for publication, acknowledges that the United States Government retains a non-exclusive, paid-up, irrevocable, worldwide license to publish or reproduce the published form of this manuscript, or allow others to do so, for United States Government purposes. The Department of Energy will provide public access to these results of federally sponsored research in accordance with the DOE Public Access Plan (http://energy.gov/downloads/doe-public-access-plan).}

\end{frontmatter}
\footnotetext[1]{\copyright 2016. This manuscript version is made available under the CC-BY-NC-ND 4.0 license http://creativecommons.org/licenses/by-nc-nd/4.0/}

\section{Introduction}

Precise determination of neutron $\beta$ decay observables allows for tests of our understanding of the electroweak interaction and searches for physics beyond the Standard Model of particle physics~\cite{Abele20081,RamseyMusolf20081, 0954-3899-36-10-104001,  RevModPhys.83.1111, ANDP:ANDP201300072, PhysRevD.88.073002, Cirigliano201393, RevModPhys.87.1483}.  There are several ongoing efforts to measure precisely the lifetime~\cite{Materne2009176, PhysRevLett.111.222501,PhysRevC.89.052501, Ezhov2014,Arimoto2015187, Arzumanov201579, PhysRevC.94.045502}, which, along with a measurement of the beta-asymmetry~\cite{1742-6596-340-1-012048, PhysRevLett.110.172502, PhysRevC.87.032501,Markisch2009216} or the $\beta$-neutrino correlation~\cite{Wietfeldt2009207, Simson2009203,Baessler2013}, can be used to test the unitarity of the Cabibbo--Kobayashi--Maskawa (CKM) quark-mixing matrix.  The neutrino asymmetry~\cite{Serebrov1998, Kreuz2005263, PhysRevLett.99.191803, Broussard2012} and the Fierz interference term (not yet directly measured in neutron decay) can be used in searches for scalar and tensor currents beyond the Standard Model~\cite{PhysRevD.85.054512,PhysRevC.87.065504,PhysRevC.88.048501,PhysRevC.92.069902,PhysRevC.89.025501,PhysRevC.91.049904}. The neutrino asymmetry is also particularly sensitive to right-handed weak currents~\cite{Yerozolimsky2000491,Abele2000499}.

Advances in detector technology are needed to improve the precision of $\beta$-decay correlations.  Recoil protons have a maximum initial energy of 751~eV and therefore require an accelerating potential to be detected.  Previous experiments in neutron beta decay have used a variety of instruments to detect the accelerated protons including CH$_4$-filled proportional counters~\cite{spivak}, photomultipliers with thin scintillating layers of CsI(TI)~\cite{EROZOLIMSKII199133} or NaI(TI)~\cite{PhysRevLett.33.41}, microchannel plates~\cite{Serebrov1998,Mostovoi2001,Beck2002}, silicon PIN diodes~\cite{Soldner200449}, conversion foils which eject secondary electrons~\cite{PhysRevD.18.3970,PhysRevLett.99.191803,PhysRevLett.100.151801,Hoedl2006}, Avalanche Photodiodes~\cite{1748-0221-11-02-C02068}, silicon drift detectors~\cite{Simson2009203}, and silicon surface barrier detectors~\cite{0295-5075-33-3-187, 0954-3899-28-6-314, PhysRevC.71.055502, PhysRevC.81.035503, Wietfeldt2009207, PhysRevC.86.035505}.  Multi-pixel silicon PIN diodes have been used for precision $\beta$ spectroscopy in tritium decay in the KATRIN experiment \cite{Amsbaugh201540}. 

The UCNB and Nab collaborations have developed a detection system which represents a significant improvement over prior approaches.  Coincident detection of electrons and recoil protons by this detector system presents a significant challenge.  General requirements for detection of protons include very low noise and a thin entrance window in addition to the accelerating potential. To accurately determine the electron energy, the detector must have sufficient thickness to fully stop the electron and a thin dead layer to limit the corrections needed for energy loss and backscattering.

UCNB and Nab have additional requirements specific to their experimental geometries.  The detectors have a large area to increase the total decay rate measured and allow for the magnetic field expansion in the spectrometer, which suppresses backscattering by longitudinalizing the charged particles' momenta.  In both experiments, backscattered electrons will be guided by the magnetic field to be detected in the same or neighboring pixel in the original detector or a corresponding pixel in the opposing detector.  Backscatters to different pixels are identified by their delayed timing relative to the original hit.  Depending on the relative signal amplitudes and signal to background, backscattering to the same pixel can be recognized with time resolution much better than the signal rise time from the shape of the waveform rising edge.  Backscattered protons do not deposit enough energy to be detected and are treated as an inefficiency.  In UCNB, identifying which detector event occurred first is necessary for determining the original direction of the electron.  Fast timing is required as the electron traverses the UCNB spectrometer in $\sim$15~ns. 

Nab also requires fast and stable timing to accurately determine the proton time of flight (TOF), with the detected electron serving as the "start" for the TOF measurement.  Furthermore, given the asymmetrical acceleration potentials applied to the Nab detectors, the bounce history of backscattered electrons must be well understood in order to correctly reconstruct the full beta energy.  In the planned follow-up experiments to Nab which utilize polarized neutrons, abBA and PANDA, fast timing is also required to determine the initial emission directions of the electron. 

In section~\ref{expov} we describe the principles of the two experiments, UCNB and Nab, and how the correlations are determined from the data.  Section~\ref{det} presents an overview of the detection system including the silicon detector, preamplifier electronics, and data acquisition.  Section~\ref{mount} reviews the mounting system which enables high voltage operation of the detector in the spectrometer.  Finally, the performance of the detection system including the demonstration of electron-proton coincidences in neutron decay is detailed in section~\ref{perf}.

\section{Experiment overview} \label{expov}

The experimental approaches for the UCNB and Nab experiments are described in Refs.~\cite{Wilburn2009, Broussard2012,0954-3899-41-11-114007} and Refs.~\cite{Bowman2005,Pocanic2009211,Baessler2013,0954-3899-41-11-114003}, respectively, and are summarized here. UCNB aims to measure the antineutrino asymmetry $B$ between the neutron spin and the emitted antineutrino momentum with better than 0.1$\%$ precision.  The experiment is designed to utilize polarized ultracold neutrons (UCN) at the Los Alamos Neutron Science Center (LANSCE) UCN facility~\cite{Saunders2013}. Nab's goal is to measure the correlation between the electron and antineutrino with precision $\Delta a/a \sim 1 \times10^{-3}$, and the Fierz interference term with precision $\Delta b\sim 3\times10^{-3}$ using an unpolarized cold neutron beam at the Fundamental Neutron Physics Beam at the Spallation Neutron Source at Oak Ridge National Laboratory.

\begin{figure}
    \centering
    \includegraphics[width=0.45\textwidth]{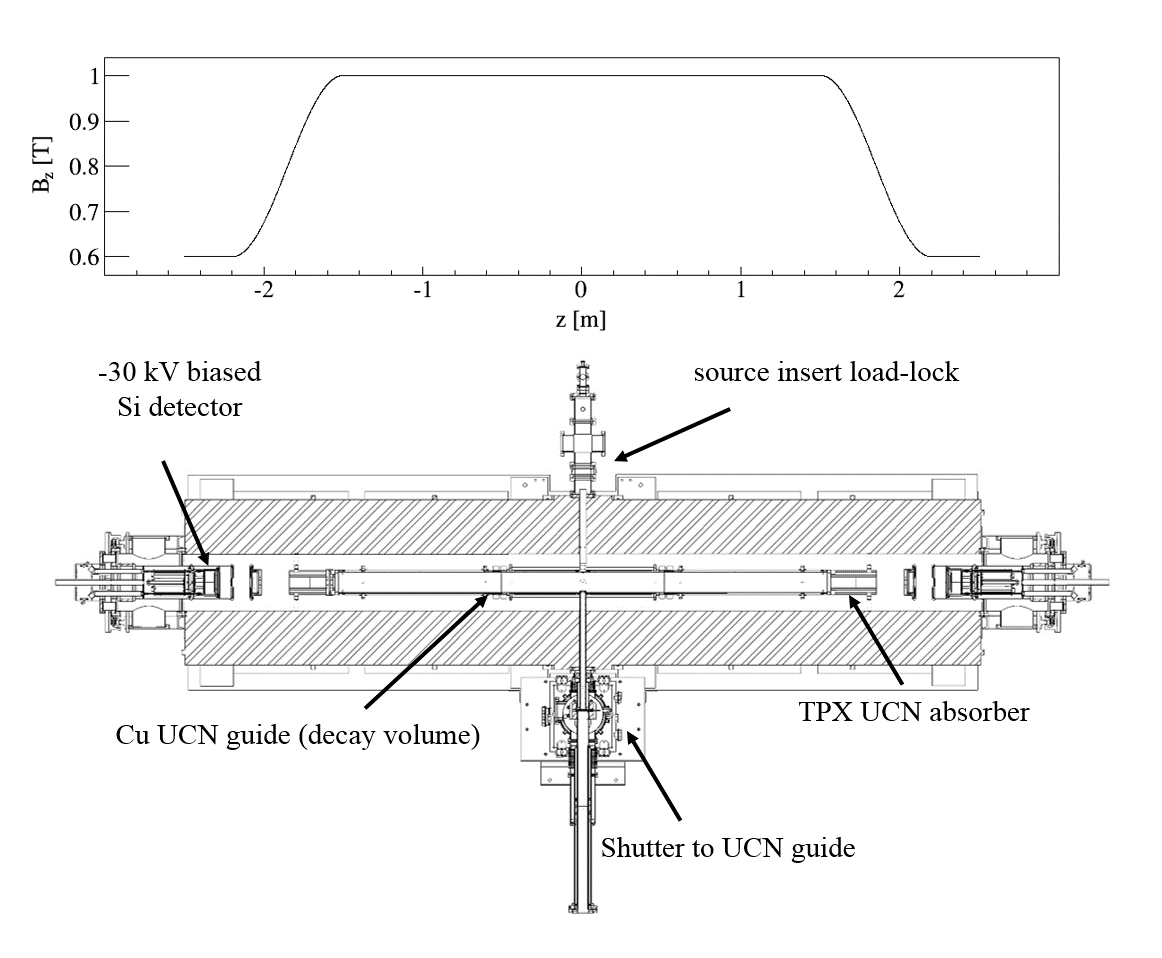}
    \caption{The UCNB experiment modifies the UCNA superconducting spectrometer (SCS) to include biasable detection systems and an open decay trap to determine the proton and electron directions. The on-axis magnetic field along the length of the spectrometer is also shown.}
    \label{fig:ucnb}
\end{figure}

In both experiments, the information about the neutrino direction/moment\-um is determined from the proton and electron directions/momenta. The UCNB experiment is performed using a modified version of the UCNA apparatus~(Fig. \ref{fig:ucnb})~\cite{PhysRevC.86.055501}. Polarized UCNs are partially bottled by the material potential of a 3~m long, 12.7~cm diameter, diamond-like carbon-coated copper guide located inside the 1~T superconducting spectrometer (SCS) magnet, with identical detectors at either end. UCNA used thin Be-coated Mylar foils at the end of the guide to completely bottle UCN while permitting electrons to pass through.  UCNB instead uses copper end caps with 3.8~cm\ diameter central holes to allow the $<$~800~eV protons that are within the viewing area of the detector to escape, while still partially bottling UCN.  A TPX Polymethylpentene~\footnote{http://www.mitsuichemicals.com/tpx.htm} tube is installed between the decay trap and the detection system.  The TPX tube absorbs UCN and reduces decays in the magnetic field expansion region.  The detectors are located in the homogeneous 0.6~T region after the magnetic field expansion.  In the final implementation of the UCNB geometry, the TPX tube will be replaced by an electrode assembly with a UCN absorbing coating to shape the electric fields.

Charged particles from the decay of the polarized UCN are guided by the magnetic field to one of the two detectors.  Using same-side coincidences, the asymmetry $B_{\text{exp}}$ can be constructed,
\begin{align}
B_{\text{exp}}(E) &= \frac{N^{--}(E) - N^{++}(E)}{N^{--}(E) + N^{++}(E)},
\end{align} 
where $N^{\beta p}$ is the number of $\beta$ particles and protons emitted in the positive or negative (relative to neutron spin) direction. This experimental asymmetry is a function of the neutrino and beta asymmeries and the beta neutrino correlation, $B$, $A$, and $a$, respectvely, and given the current precision of the beta asymmetry and beta neutrino correlation, can be used to extract $B$ at the 0.1\% level~\cite{PhysRevLett.99.191803}.​  Both detection systems are biased to $-30$~kV to accelerate the protons with enough energy to be detected, while the UCN decay trap is electrically grounded.  A modified version of the UCNB geometry dedicated to measurements of the spin-averaged electron spectrum is also under development.

\begin{figure}[h]
    \centering
    \includegraphics[width=0.35\textwidth]{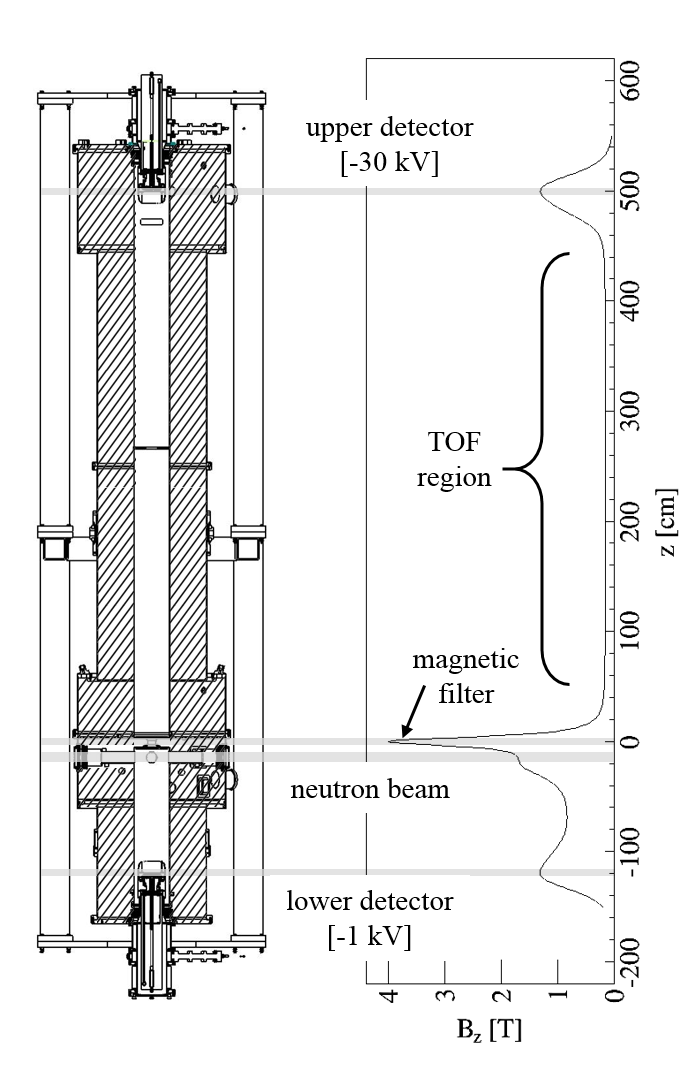}
    \caption{The Nab spectrometer utilizes an asymmetric configuration to determine the proton momentum from the time of flight to the upper detector and the electron energy in either detector, with a magnetic filter that permits only upward-going protons. The on-axis magnetic field along the length of the spectrometer is also shown.}
    \label{fig:nab}
\end{figure}

The Nab spectrometer has an asymmetric magnetic field configuration with respect to the decay volume, with a short and long flight path from the cold neutron beam at 2~T to the two detection systems placed at 1~T after the field expansion~(Fig. \ref{fig:nab}). The magnetic filter before the long flight path has a peak magnetic field of 4~T to accept only protons with small angle between their momentum and the magnetic field.  Thanks to conservation of momentum, the angle $\theta_{e \nu}$ between the electron and the antineutrino depends only on the proton momentum $p_p$ and the electron energy $E_e$, and to a good approximation the electron and antineutrino momenta $p_e$ and $p_\nu$ depend only on the electron energy. The correlation $a$ can therefore be taken from the slope in the proton momentum distribution $P(p_p^2)$,
\begin{align}
P(p_p^2) &\propto 1 + a \frac{p_e}{E_e}cos\theta_{e \nu}, \\
cos \theta_{e \nu} &= \frac{p_p^2  -  p_e^2 -  p_\nu^2}{2 p_e p_\nu}
\end{align}

The detector after the long flight path will be biased to $-30$~kV to detect protons, while the short flight path detector will be biased to $-1$~kV to suppress reflections of downward-going protons from the lower detector.  The proton momentum will be determined from the time of flight $t_p$ across the 5~m distance, which uses the magnetic field expansion to longitudinalize the momentum, and electrons will be detected in both detectors. The $a$ correlation is then determined from the slopes of the $t_p^{-2}$ distributions at constant $\beta$ energy.  An alternate configuration will be used for a measurement of the Fierz term, $b$.  As the proton momentum is not needed, the short flight path detector will be biased to $-30$~kV and the long flight path detector biased to $+1$~kV to reflect upward-going protons back to the lower detector, to improve the measured decay rate.

\section{Detection system}\label{det}

At the heart of the detection system are 1.5 to 2~mm thick, highly segmented silicon detectors, with active area diameter of 11.5~cm, and thin front dead layer, developed in collaboration with Micron Semiconductor, Ltd.~\cite{Micron} The detector is an n-type on n-type design.  The front junction for the active area is created by a shallow boron implant to give a reduced dead layer, while the periphery has a deep implant to ensure good contact (see Fig.~2 in Ref.~\cite{SalasBacci2014408}).  The junction face is metallized with a 300~nm thick, square aluminum grid to improve charge collection.  The grid covers 0.4\% of the surface area of the detector.  The ohmic side is created by a phosphorus implant and is divided into 127~hexagonal pixels of 70 mm$^2$ area plus an outer ring of partial pixels. The contacts are hexagonal pads separated by 100~$\mu$m gaps, so that there is no dead space between pixels, and all charge is fully collected. The ohmic side is mounted directly to a ceramic interface with ultrasonic wire bonded contacts to each of the pixel faces and to the junction-side detector bias and guard rings.

Two detectors, one 0.5~mm and one 1~mm thick, were characterized using the central detector pixel at the proton accelerator at Triangle Universities Nuclear Laboratory as described in Ref.~\cite{SalasBacci2014408} and key results are summarized here. The rectifying junction on the front face of the detector creates a dead layer through which charged particles must pass to be detected. This was measured to be $\sim$100~nm silicon equivalent using a proton and deuteron beam.  A noise threshold of 6~keV and energy resolution of 3~keV FWHM were measured.  Protons were distinguished from noise with accelerating voltages as low as 15~kV.  The detected energy after the deadlayer for these protons was about 9~keV.  The detector exceeds the requirements for detecting protons accelerated to 30~keV, which deposit $<$20~keV of energy after traversing the dead layer.

\begin{figure}
    \centering
    \includegraphics[width=0.35\textwidth]{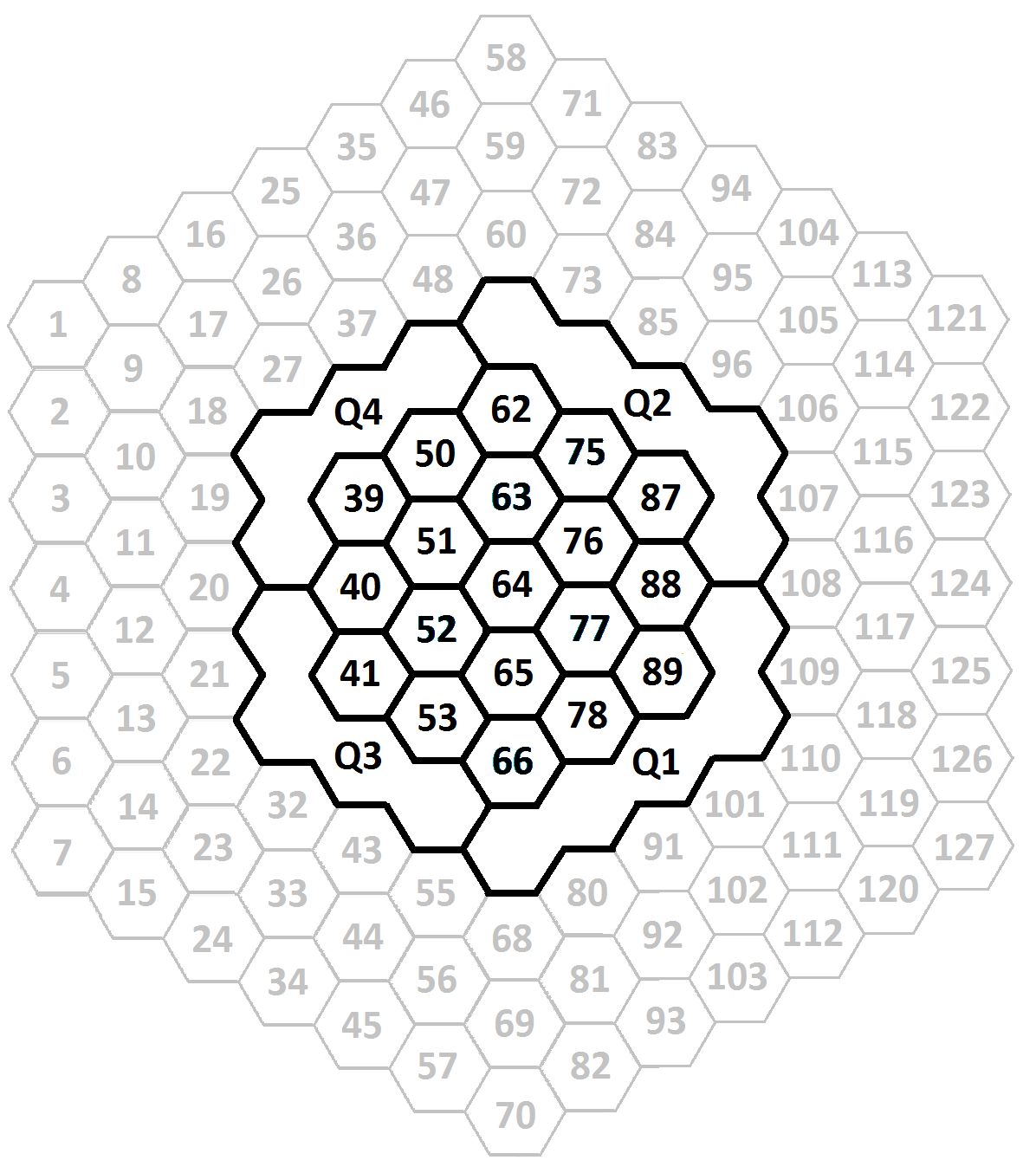}
    \caption{The silicon detector was instrumented with central 19 individual pixels and outer ring of 18 pixels ganged into 4 channels.}
    \label{fig:pixel}
\end{figure}

Subsequently, a high gain, compact, 24~channel preamplification system with 20~ns timing  was developed to read the central 19 pixels, plus 4 channels of ganged pixels and one channel reserved for a pulser input (Fig. \ref{fig:pixel}).  The ganged pixels suffered from too great a capacitance mismatch with the electronics and were too noisy to be useful for proton detection.  The effect of the capacitance mismatch was later confirmed qualitatively using a pulser circuit input to the electronics with the detector represented by a capacitor.  The preamplifier assembly was based on the electronics chain described in Ref.~\cite{SalasBacci2014408}. The assembly is compact due to the size constraints for installation in the spectrometer.

\begin{figure}
    \centering
    \includegraphics[width=0.45\textwidth]{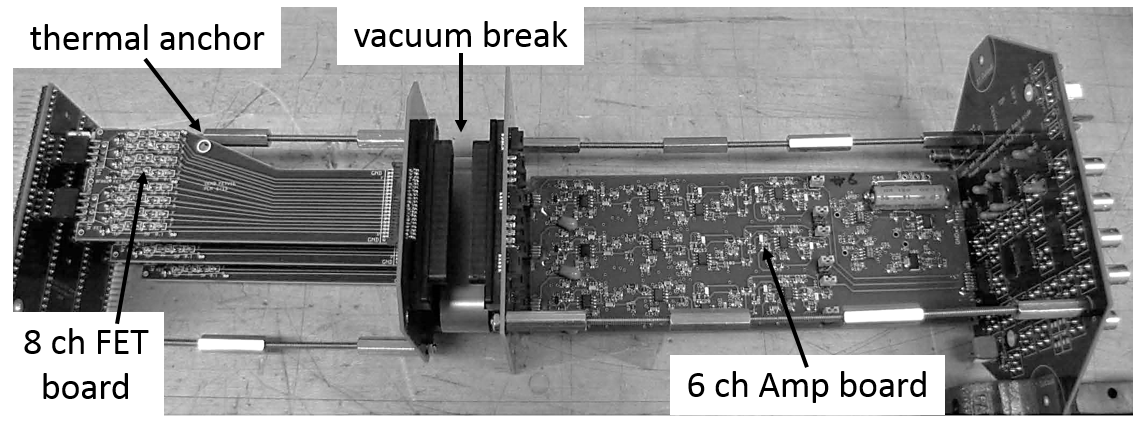}
    \caption{The preamplifier includes a FET subsystem mounted in vacuum and an amplifier subsystem mounted in air.}
    \label{fig:preamp}
\end{figure}

The preamplifier is divided into two subsystems (Fig.~\ref{fig:preamp}).  The FET subsystem contains the low noise FETs and feedback loop and resides in vacuum immediately behind the detector.  It consists of 3 parallel boards of 8 channels which mount to the detector through plastic multi-pin socket connectors.  For the fully instrumented geometry, the frontend electronics will instead be connected to the detector through pogo-pin connectors, similar to the KATRIN scheme~\cite{Amsbaugh201540}. The FET volume is housed between the $\sim$100~K detector and the room temperature feedthrough to the second subsystem, requiring long ($\sim$11.5~cm) FET boards to accommodate the large temperature gradient.  To improve cooling of the BF862 n-JFETs, the FET board is thermally anchored to the liquid nitrogen cooled copper can surrounding the assembly, which cools the detector.

\begin{figure}
    \centering
    \includegraphics[width=0.45\textwidth]{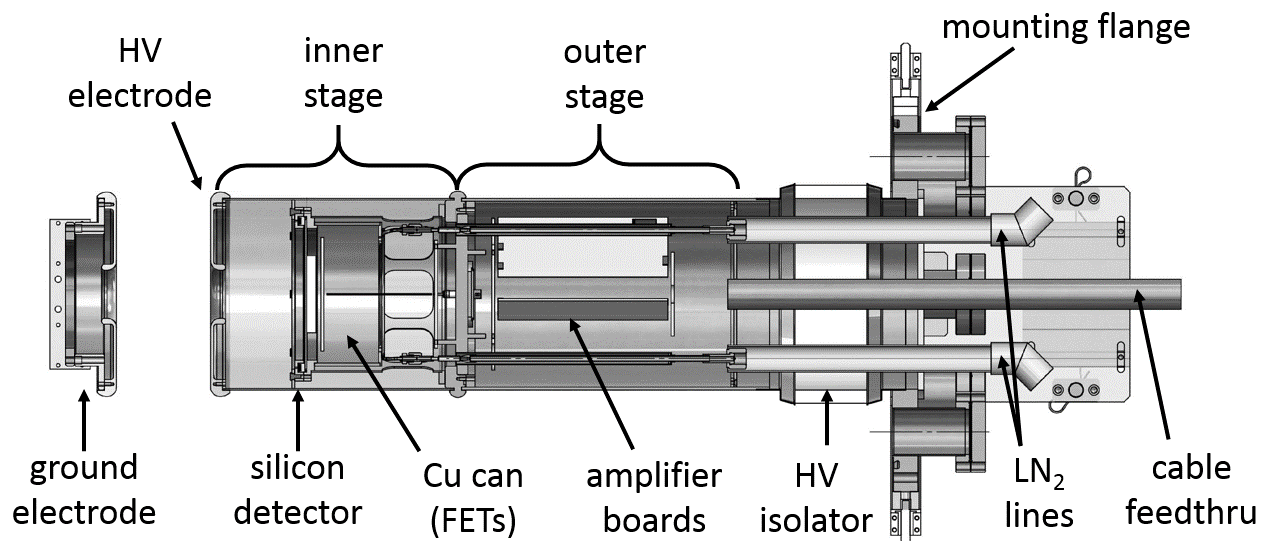}
    \caption{The detector mount houses the detector, preamplifier electronics, liquid nitrogen lines, and allows for high voltage bias up to 30~kV.  The inner stage is in vacuum and the outer stage is in air.}
    \label{fig:mount}
\end{figure}

The amplifier subsystem contains the later gain stages and resides in air.  The temperature of this subsystem is maintained at room temperature by forced dry nitrogen gas flow.  It consists of 4 parallel boards of 6 channels. Each channel is integrated by an AD8011 op-amp based circuit followed by two stages of low-noise, low-distortion AD8099 op-amps, for a total amplifier gain of 80~mV/fC.  The preamp saturates at voltages corresponding to about 600~keV energy deposition, below the $\beta$ endpoint.  This gain setting is higher than that planned for the actual experiment, but was selected to improve the discrimination of $<20$~keV protons.  The outputs can be taken as fast signals or passed (jumper selectable) to a shaping circuit. Improvements to the electronics from Ref.~\cite{SalasBacci2014408} include filtering on the final amplifier stage to reduce pickup, reduced negative feedback in the first amplifier stage to improve the rise time, increased FET drain resistance/voltage to reduce Johnson noise, and low dropout regulators to reduce power consumption.  The preamplifier is powered by  $+12$~V and $\pm$6~V outputs from a Keysight N6700B mainframe with a N6733B 20V 2.5A and two N6732B 8V 6.25A power modules, respectively.

The preamplifier outputs drive 50~Ohm single-ended signals over coaxial cables to the data acquisition system (DAQ) inputs. The DAQ was based on the National Instruments 250~MHz 14~bit PXIe-5171R digitizers~\cite{NI} using custom FPGA firmware developed specifically for this application.  The NI PXIe-8135 controller acted as the DAQ computer, installed in the NI PXIe-1062Q chassis, which, in turn, was housed inside the high voltage safety enclosure.  Fiber optic links were used to communicate with the biased components.

\section{Mounting System}\label{mount}

The prototype detector mount assembly (Fig.~\ref{fig:mount}) ensures stable and consistent operation at $-30$~kV bias and magnetic field of 0.6~T, and limits Penning traps to minimize possibility of damage to the detector or electronics from electrical breakdown. It consists of a 16.5~in.\ conflat flange which mounts directly to either end of the SCS. The outer flange adapts to a re-entrant, 9.1~in. O.D., $55$~kV rated CeramTec isolator brazed to Kovar, which is welded to a two-stage stainless steel tube. The length was selected to position the detector in the magnetic field expansion region of the UCNB spectrometer. The outer stage contains electrical components that are maintained at room temperature and pressure, and is separated by a vacuum feedthrough from the inner stage. Inside the inner stage is the concentric copper can that is hard-soldered to liquid nitrogen cooling lines that feed through from the outer stage. This section includes all electronics components which reside in the main spectrometer vacuum.  The silicon detector's ceramic interface is mounted at the front face of the copper can and is cooled directly by it.

A stainless steel cover surrounds the inner stage to provide the polished surface of the high voltage electrode. The front face of the cover has a 5~cm diameter window for charged particles to reach the detector. A ground electrode is situated 10~cm upstream, between the mount and the UCN decay trap.  A TPX tube between the ground electrode and the decay trap increases the chance that UCNs that escape the decay trap are absorbed and do not decay in the magnetic field expansion region.  The detector mount beyond the ceramic isolator is maintained at -30~kV high voltage during operation. The nitrogen cooling lines are vacuum jacketed through both stages, and separated by a G10 high voltage break at the ceramic isolator region from the lines to the liquid nitrogen dewar. 

The system is cooled to reduce the detector leakage current, and to reduce the FET noise.  The effect of temperature on the leakage current is discussed in Section~\ref{leak}.  The cooling requirement was investigated using Lakeshore DT-670 calibrated silicon diode sensors.  Outside of the vacuum, the heat load from the preamplifier electronics is about 5~W per board.  The outside section is maintained near room temperature by flowing cool, dry nitrogen gas during operation.  The heat load from the FET assembly was measured to be about 200~mW per board. From finite element simulations, the heat load conducted through the copper ground plane of the FET cards from the room temperature feedthroughs is estimated to be the dominant contribution, of about 2~W per board, or 6~W total for the prototype, and an expected 30~W for the fully instrumented system.

In the Nab spectrometer, the detector will be surround\-ed by a cooled electrode, and surfaces outside with direct view on the detector will be at least as cold as the detector. The detector as installed in the prototype mount for this cooling study instead viewed the room temperature test chamber and cooled electrode cap.  The ceramic backing of the detector reached an equilibrium temperature of about 175~K at the detector center during normal operation, with a 5~K temperature gradient to the (cooled) detector edge.  The FET itself reached an ultimate temperature of 185~K.  The temperatures of the FET board at the mid point (near thermal anchor) and feedthrough (warm side) reached 200~K and 220~K, respectively.  From observation of the noise floor as cooling to the detection system is halted, it is expected that performance of the detection system should improve further with colder operation.  The fully instrumented system is designed to achieve a detector temperature of 110~K.

The stability of high voltage was significantly improved by surrounding the biased part of the electrode up to the ceramic isolator with a Mylar sheet, thus preventing electrical breakdown at the ceramic weld. Moisture buildup on the G10 nitrogen lines during detector cooling cycles was one of the primary remaining causes of high voltage breakdown. This was alleviated by surrounding those sections with larger diameter G10 tubes and blowing dry nitrogen gas through them. The high voltage assembly has maintained stable operation with magnetic fields up to 0.6~T at the detector and $-30$~kV bias with no high voltage-related damage to the detector or electronics.

Electronics cabling and power were routed out of the detector mount through a 5~cm diameter copper tube inside a thick-walled PVC pipe to a separate high voltage safety box containing the DAQ.  The copper tube was electrically connected to the electrode assembly beyond the ceramic isolator and the interior of the high voltage safety box.   The high voltage was supplied by a Matsusada Precision EQ-30N1 $-30$~kV supply through a Cableform CJG 100~MOhm resistor to the copper tube. The DAQ box consisted of an outer aluminum frame lined internally with 1.25~cm of polyethylene and polycarbonate sheets, with 1.25~cm diameter vent holes in the top and bottom plates for cooling using external fans. The sheets were fastened to the outer frame with nylon bolts covered by extra layer of 1.25~cm thick polyethylene.  A Stangenes 1~kVA, 50~kV isolation transformer provided power to components inside the box.  The electronics resided within an internal frame of 5~cm diameter copper piping for protection from high voltage arcing.  

The DAQ box also served the function of an installation cart for the detector mount, and was mounted to the spectrometer during operation.  The magnetic field at the location of the DAQ box ranged from 0.01--0.1~T when the spectrometer was operated at full field, which caused occasional performance problems with some hardware.  An independent DAQ box will be developed to move the electronics further outside the field in Nab and UCNB.

\section{Detector Performance}\label{perf}

The detection system characteristics were evaluated using pulsers and sealed conversion electron and X-ray sources.  Electron-proton coincidences were measured from UCN $\beta$ decay.  The 1.5 mm thick detectors were used in the measurements reported here.

\subsection{Rise time}

\begin{figure}
    \centering
    \includegraphics[width=0.45\textwidth]{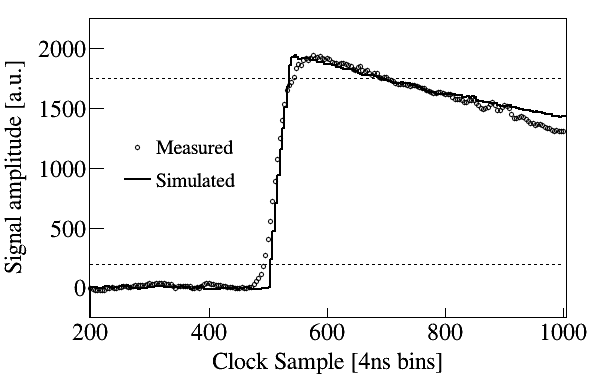}
    \caption{Waveform of a 363.8~keV conversion electron from $^{113}$Sn with rise time of 44~ns (50~ns average) and simulated detector signal waveform.  Dashed horizontal lines mark 10$\%$ and 90$\%$ of the maximum amplitude.}
    \label{fig:risetime}
\end{figure}

The signal rise time of the preamplifier electronics alone, defined as the time for the waveform amplitude to change from 10\% to 90\%, was measured to be 20~ns under ideal, test-bench conditions.  The average rise time of signals from the full detection system was measured to be $\sim$50~ns using conversion electron sources.  The measured rise time agrees reasonably with a pulse rise time estimated by modeling the charge collection time in the silicon detector using a parallel plate geometry~\cite{spieler2005}, which depends on the detector thickness/capacitance, bias voltage, and temperature (through the electron/hole mobilities~\cite{knoll}) (Fig.~\ref{fig:risetime}).
 
\subsection{Linearity}

The silicon detector is expected to be highly linear~\cite{PEHL196845}.  The linearity of the preamplifier electronics and data acquisition was measured by sending a square wave pulse through a high pass filter into the FET board at room temperature.  For each input amplitude, 10$^6$ output waveforms were collected.  The waveforms were shifted to a common start time defined from the rising edge (to within a tenth of a clock tick), resulting in a distribution of pulse shapes.  An average waveform for each input amplitude was produced by taking the mean of a fit of the distribution at each partial clock tick to a Gaussian.  Slight variations in the pulse shape were observed for each amplitude, due to ringing from the square wave pulser.

To extract the pulse amplitude more accurately, the average waveform from each set was normalized by amplitude as determined from a fit to a semi-gaussian CR-RC$^n$ shaper~\cite{spieler2005}, 
\begin{align}
V(t) = \left(\frac{t}{\tau}\right)^n e^{\frac{-t}{\tau}},
\end{align}
where $V(t)$ is the amplitude at sample $t$, $\tau$ is the shaping time, and $n$ is the number of integrators.  The set of averages obtained after normalization were subsequently summed and averaged to produce a single average pulse shape.  Waveforms beyond the saturation point for the preamp were not included in this average.  The average waveforms for each amplitude were then fit to this pulse shape.  The pulses appeared to be uniform with amplitude:  the maximum residuals from the fit to the average for each amplitude were up to 2--3$\%$ due to the ringing effect and $<1\%$ otherwise.

\begin{figure}
    \centering
    \includegraphics[width=0.45\textwidth]{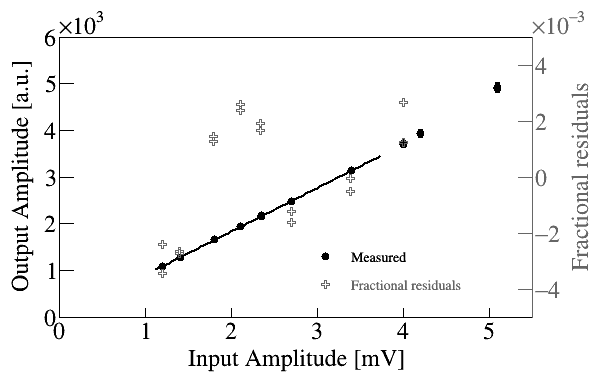}
    \caption{Measured linearity of preamplifier electronics and DAQ.  The fitted region corresponds to amplitudes below the saturation point of the preamplifier.  The maximum fractional nonlinearity observed was 0.3\%.}
    \label{fig:lin}
\end{figure}

The extracted amplitudes are plotted against the corresponding input amplitude of the function generator in Fig.~\ref{fig:lin}.  Two data sets were taken at each input amplitude to explore the uncertainty due to the ringing.  The nonlinearity is less than 0.3$\%$ for amplitudes below preamp saturation.  Beyond the saturation point, the nonlinearity was as high as 3.5$\%$, due to poor fit of the distorted pulse shapes.  The linearity of the fully instrumented system will ultimately be characterized using calibration sources.

\subsection{Electronic noise analysis}

\begin{figure}
    \centering
    \includegraphics[width=0.45\textwidth]{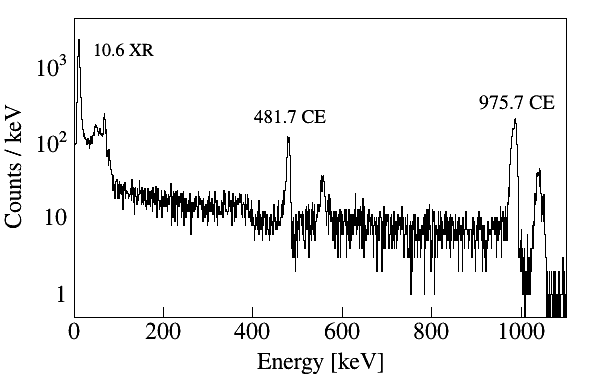}
    \caption{ Conversion electron and X-ray energy spectrum of $^{207}$Bi.  At 10.6 keV is the X-ray emission line, at 481.7 keV is the K-shell emission line followed by L- and M-shell emission lines at 553.8 and 565.8 keV, respectively. The Compton edge from the 570~keV gamma-ray is visible at 390~keV. The 975.6keV K-shell emission line and 1047.8 and 1059.8 keV L- and M-shell lines are beyond the saturation point of the preamplifier, and their measured energies are distorted.
}
    \label{fig:biplot}
\end{figure}

The detected energy threshold was determined to be about 8~keV for most channels.  The 10.6~keV X-ray from $^{207}$Bi can be clearly seen in the energy spectrum shown in Fig.~\ref{fig:biplot}, where the amplitude is reduced due to the trigger efficiency.  The threshold is not represented as a hard cutoff due to the imperfect filter response at low signal to noise, resulting in a gradual decrease in trigger efficiency.  The trigger efficiency as a function of energy of the final detection system will be studied using a proton source.

The sensitivity to noise was studied by extracting the energy resolution of line sources as a function of the shaping time $\tau_S$ (rise time) of a trapezoidal filter~\cite{JORDANOV1994261}.  The equivalent noise charge (ENC) is related to the energy resolution full width half maximum (FWHM) by the hole-pair liberation energy ($\sim$3.7~eV/e for the cooled detector),
\begin{align}
\text{ENC [e]} &=  \frac{\text{FWHM [eV]}}{2.35 \times 3.7 \left[\frac{eV}{e}\right]}.
\end{align}
The noise contributions include series voltage noise ($\propto$ 1/$\tau_S$) due to the FET and amplifier noise and in-series resistance, parallel current noise ($\propto$ $\tau_S$) due to the FET shot noise and detector leakage current and thermal noise from the feedback resistor, and 1/$f$ noise from various components (independent of $\tau_S$)~\cite{Bertuccio93},
\begin{align}
\text{ENC}^2 [e^2] &=  \frac{h_1 [e^2 \mu s]}{\tau_S [\mu s]} + h_2 [e^2]+ h_3 \left[\frac{e^2}{\mu s}\right] \tau_S [\mu s]. \label{eq:enc}
\end{align}

The monoenergetic conversion electron and X-ray sour\-ces used to study electronic noise included $^{207}$Bi, $^{139}$Ce, and $^{113}$Sn, with emission line energies used in this analysis tabulated in Table \ref{tab:source}.  These energies are relevant to neutron $\beta$ decay, ranging from $\sim$10~keV to $\sim$1~MeV.  The sources were installed directly in front of the detector to expose the detector to the conversion electrons and X-rays.  The sources were also installed in the center of the spectrometer (2~m away) through the load-lock insert, to view only the electrons, which are guided by the magnetic field.  The sources were sealed between aluminized Mylar foils, resulting in some energy loss of the conversion electrons. The foil thickness was measured to be 9.5~$\mu$m~\cite{PhysRevC.87.032501} and the energy loss in the foil and deposited energy in the detector was simulated using PENELOPE~\cite{salvat2006penelope}.

\begin{table}
\centering
  \caption {Mono-energetic sources (XR = X-rays and CE = conversion electrons) and the emission line energies used to characterize the detection system~\cite{NNDC}.}
  \label{tab:source}
    \begin{tabular}{cccc}
\hline
      Isotope & Energy (keV) & Intensity ($\%$) & Source \\
      \hline
      $^{207}$Bi & 10.6 & 33.2 & XR \\
      $^{113}$Sn & 24.0 & 28.0 & XR \\
      $^{113}$Sn & 24.2 & 51.8 & XR \\
      $^{113}$Sn & 27.2 & 4.7 & XR \\
      $^{113}$Sn & 27.3 & 9.0 & XR \\
      $^{113}$Sn & 27.9 & 2.39 & XR \\
      $^{139}$Ce & 33.0 & 22.5 & XR \\
      $^{139}$Ce & 33.4 & 41.0 & XR \\
      $^{139}$Ce & 37.7 & 4.0 & XR \\
      $^{139}$Ce & 37.8 & 7.6 & XR \\
      $^{139}$Ce & 38.7 & 2.5 & XR \\
      $^{113}$Sn & 363.8 & 28.8 & CE \\
      $^{113}$Sn & 387.5 & 5.6 & CE \\
      $^{113}$Sn & 390.9 & 1.1 & CE \\
      $^{207}$Bi & 481.7 & 1.5 & CE \\
      $^{207}$Bi & 553.8 & 0.4 & CE \\
      $^{207}$Bi & 565.8 & 0.1 & CE \\
\hline
    \end{tabular}
\end{table}

\begin{figure}
    \centering
    \includegraphics[width=0.45\textwidth]{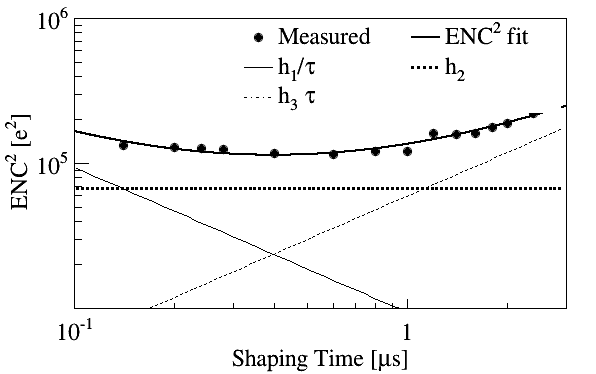}
    \caption{Squared equivalent noise charge from a typical pixel as a function of shaping time of a trapezoidal filter, extracted from resolution of $^{113}$Sn X-rays.  The fit to Eq.~\ref{eq:enc} and the individual terms $h_i$ are also shown.}
    \label{fig:shaping}
\end{figure}

Energy spectra were produced for each value of the trapezoidal shaping time.  Source lines of similar energy were fit to multiple Gaussians with a common width.  A typical shaping time scan for a single detector pixel is shown in Fig.~\ref{fig:shaping}. We expect, in general, for the line width to include an energy independent contribution from the noise and energy dependent contributions from the number of charge carriers and charge collection effects~\cite{knoll}, as well as energy loss in the source foils. The data suggests that contributions from charge collection effects are well below the level of electronic noise.  The increase in line width from the source foil was confirmed by simulation. 

\begin{figure}
    \centering
    \includegraphics[width=0.45\textwidth]{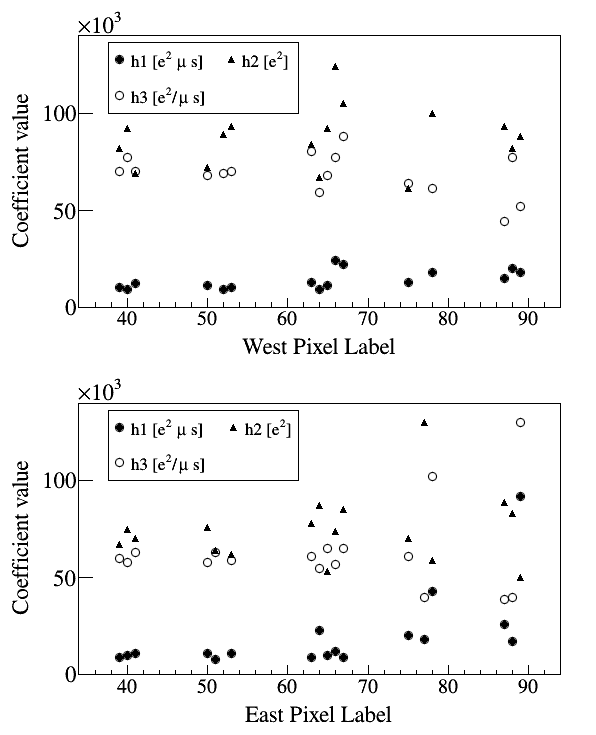}
    \caption{Noise coefficients from Eq.~\ref{eq:enc} for two 1.5 mm detectors, designated West (top) and East (bottom) for their location in the spectrometer, extracted using $^{113}$Sn X-rays or $^{139}$Ce X-rays (see Table~\ref{tab:shapescan}).}
    \label{fig:shapetable}
\end{figure}

\begin{figure}
    \centering
    \includegraphics[width=0.45\textwidth]{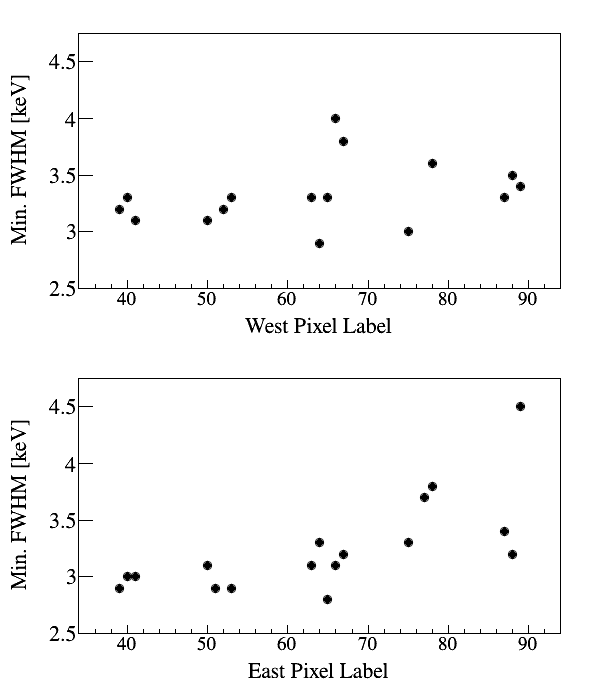}
    \caption{Energy resolution (FWHM) per pixel extracted from the minimum of the fit to Eq.~\ref{eq:enc} at the optimum shaping time for this configuration of 0.4~$\mu$s.}
    \label{fig:FWHM}
\end{figure}

The noise coefficients per pixel are tabulated in Table~\ref{tab:shapescan} and are plotted in Fig.~\ref{fig:shapetable}. The coefficients are taken from a shaping time scan using the $^{113}$Sn X-rays or the $^{139}$Ce X-rays (indicated with asterisks).  Some preamplifier channels were sensitive to microphonics from the LN$_2$ cooling or sources of parallel noise on the ground.  One possible reason is that these channels had exposed traces in view of the bias cable, which could not be completely shielded at the connection to the detector.  These channels were too noisy to perform a complete shaping scan with the $^{113}$Sn X-rays. Some pixels are omitted due to unusable electronics channels, due to damaged components.  The FWHM obtained from the minima of the fitted curves is plotted in Fig.~\ref{fig:FWHM}.  The optimum shaping time for this configuration was 0.40~$\mu$s on average for all channels, with a standard deviation of 0.03~$\mu$s.

The coefficients obtained from the fit to Eq.~\ref{eq:enc} can be used as a measure of the uniformity of the pixels and electronics channels.  The $h_1$ term is proportional to the squared total capacitance of the pixel, the FET input capacitance, the feedback capacitance, and stray capacitances.  The stray capacitance varies from channel to channel, as the traces are not laid out uniformly and the lengths and geometry vary.  The stray capacitance is of the same order as the pixel capacitance, and the variation can be attributed in part to both.  The utilization of pogo-pin connectors in the fully instrumented geometry is expected to result in a more uniform geometry and reduced stray capacitance.  The standard deviation of $\sqrt{h_1}$ is about 7$\%$ for scans performed with $^{113}$Sn, and about 30$\%$ for all pixels.  The $h_3$ term is sensitive to the leakage current, which can vary depending on the concentration of imperfections and impurities in the crystal.  However, as shown in the next section, other sources of parallel noise sources dominate.

The primary impact of the loss of pixels due to noise is the reduced fiducial volume.  In the event of a dead or unusable noisy pixel, coincidences cannot be formed in the pixel or its neighboring pixels.  Next-to-nearest neighboring pixels may also need to be included.  These pixels would be treated as a hole in the detector.  A single dead pixel plus neighboring pixels would result in a $\sim$10$\%$ or more reduction in rate, and more than two dead pixels would necessitate the replacement of the detector, depending on location.  If a pixel is usable for electrons but too noisy to detect protons, no coincidences would be formed using that pixel, only.  Generally speaking, high noise sensitivity primarily affects proton detection efficiency, as the energy resolution for the electron is only needed for a measurement of the Fierz term.

\begin{figure}
    \centering
    \includegraphics[width=0.45\textwidth]{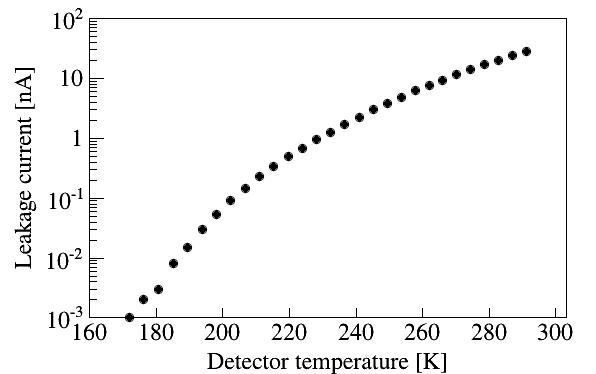}
    \caption{Leakage current of the central detector pixel as a function of detector temperature.}
    \label{fig:leak}
\end{figure}

\subsection{Leakage current}\label{leak}

The detector leakage current per pixel was measured as a function of temperature (Fig.~\ref{fig:leak}).  The total detector leakage current was also measured through the detector bias when installed in the SCS, and was typically 0.1--0.2 nA at 175~K.  Assuming the leakage current is divided among the pixels and partial pixels evenly and that the bulk contribution dominates, the leakage current per pixel was $\sim$1~pA at 175~K, in agreement with Fig.~\ref{fig:leak}. 

\begin{figure}
    \centering
    \includegraphics[width=0.45\textwidth]{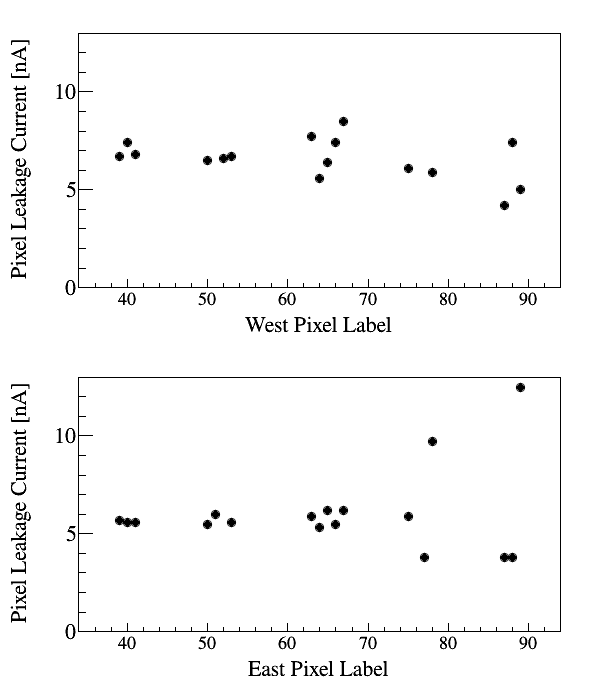}
    \caption{Leakage current per pixel extracted from the noise coefficient $h_3$ in Eq.~\ref{eq:h3}, under the (incorrect) assumption that the leakage current dominates.}
    \label{fig:I_L}
\end{figure}

In the limit that the leakage current $I_L$ per pixel is the dominant contribution to parallel noise, it can be extracted from the shaping time scan,
\begin{align}
h_3 \left[\frac{e^2}{\mu s}\right] &\approx \frac{q[e] I_L \left[\frac{e}{\mu s}\right]}{A}, \label{eq:h3}
\end{align}
where $q$ is the electron charge and $A=1.67$ is a weighting factor dependent on the trapezoidal shaping response.  However, the $I_L$ extracted from the shaping time corresponds to several nA per pixel (Fig.~\ref{fig:I_L}), significantly higher than the measured value in Fig.~\ref{fig:leak}.  We therefore understand that leakage current does not provide the dominant contribution to our parallel noise.  Based on measurements of our high density front end and preamp on a test bench and in the spectrometer, we see contributions from the leakage current to the resolution are negligible below about 175 K.  Other sources of noise, however, such as noise on the bias filter ground relative to the FET grounds, were a potential source of parallel noise which could easily account for the entire parallel noise budget in our current implementation of the detector mount and grounds in the spectrometer.  

The pixels with greater sensitivity to noise correspond\-ed to those with exposed traces on the underside of the preamplifier interface board, facing the detector. Therefore the large variation is attributed to greater coupling with the partly exposed bias line.  This effect is expected to be reduced by a redesigned interface to the detector, including a completely shielded bias connection.  Preliminary tests of cooling using cold helium gas instead of LN$_2$ also show promising results with significant reduction of microphonic noise.

\subsection{Neutron $\beta$ decay}

Coincident detection of protons and electrons from neutron $\beta$ decay provides an important validation of the detection system performance.  A series of measurements were performed using the LANSCE UCN source to demonstrate this capability.  Before entering the spectrometer the UCN passed through a 6~T magnet for nearly 100$\%$ polarization, followed by an adiabatic fast passage spin flipper~\cite{Holley12}.  The ultracold neutron rates throughout the neutron guide system were monitored using $^3$He-filled multiwire proportional chambers~\cite{Morris2009248} and $^{10}$B-based multilayer surface detectors~\cite{Wang201530}.  For this measurement only one detector was biased to -30~kV.

\begin{figure}
    \centering
    \includegraphics[width=0.45\textwidth]{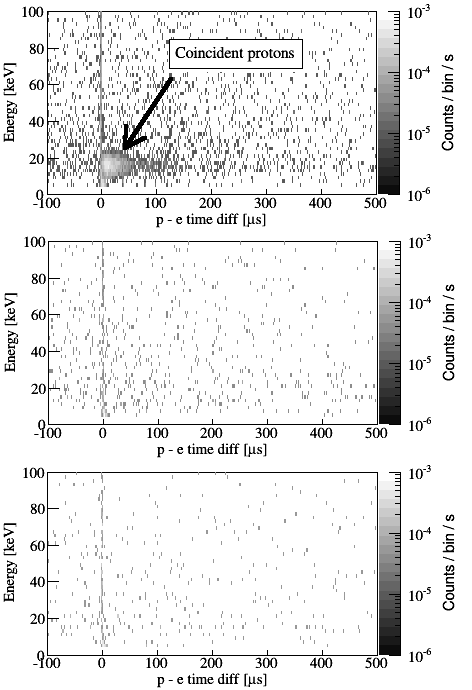}
    \caption{Event energy vs.\ time since previous event for same-side coincidences summed over all pixels from polarized UCN $\beta$ decay.  Protons are only observed when an accelerating voltage is applied and UCN are in the decay volume (top).  The background rate from UCN but no accelerating voltage (middle) and accelerating voltage but no UCN (bottom) is also shown.}
    \label{fig:protons}
\end{figure}

UCN $\beta$ decay data were taken in a sequence of $\beta$ decay measurements, background measurements, and depolarization measurements for each neutron spin state (Table~\ref{tab:sequence}).  Energy vs.\ timing spectra for same-side coincidences were compared with and without UCN in the apparatus, and with and without accelerating voltage applied (Fig. \ref{fig:protons}).  A peak corresponding to protons only appeared when both conditions were true.  Valid proton events had energy~$\lesssim$~20~keV, and most arrived within a few tens of $\mu$s after an electron; longer times correspond to larger pitch angles.  The events with time difference near zero correspond to backscattered electrons.  An average rate of $3\times10^{-3}$~same-side coincidences/pixel/s averaged over both spin states was measured.  A background rate of $3\times10^{-4}$ coincidences/pixel/s was measured.  A total of 27 hours of $\beta$ decay data with HV on, 7.5 hours of $\beta$ decay data with HV off, and 4.5 hours of background data were taken.

\begin{table}
  \caption {Run sequence for UCN $\beta$ decay measurements included normal data collection for both spin states ($\beta$), data collection with high voltage (HV) off to collect only $\beta$ particles ($\beta$*), in situ measurements of depolarization  in the decay trap (D), and background  measurements with no UCN loaded into the decay trap (B).}
\centering
  \label{tab:sequence}
    \begin{tabular}{ccc}
\hline
      Type & HV & Spin \\
      \hline
      $\beta$ & on & + \\
      $\beta$* & off & + \\
      $\beta$ & on & + \\
      D & on & +~$\rightarrow$~- \\
      B & on & N/A \\
      $\beta$ & on & - \\
      $\beta$* & off & - \\
      $\beta$ & on & - \\
      D & on & -~$\rightarrow$~+ \\
      B & on & N/A \\
\hline
    \end{tabular}
\end{table}

\begin{figure}
    \centering
    \includegraphics[width=0.45\textwidth]{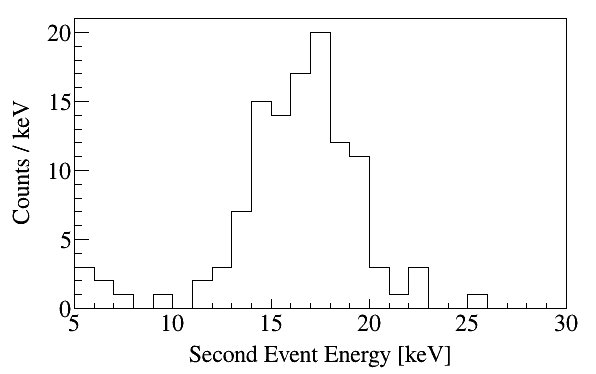}
    \caption{Energy of second events in coincidence window between 3~$\mu$s and 500~$\mu$s for a typical measurement (45~min).  In this run, a peak corresponding to protons appears at 16.5~keV, which is consistent with 50~nm water layer adsorbed on the detector surface.}
    \label{fig:esum}
\end{figure}

The energies of the second events in the coincidence window for a typical measurement are plotted in Fig.~\ref{fig:esum}.  ​The distribution of protons is clearly separated from the noise pedestal, although trigger efficiency for lower energy protons needs further study.  The proton peaks were observed at maximum energy of 19~keV and $\sim$5~keV~FWHM, as expected from a GEANT4~\cite{Agostinelli2003250} and SRIM~\cite{ZiegetalSRIM2010} simulation of 30~keV protons through a 100~nm silicon dead layer.   The peak was observed to reduce in energy by 0.5~keV/h, resetting when the detector was warmed.  This may be attributed to residual water vapor adsorbing onto the detector surface at a rate of 5~nm thickness per hour.  The inadequate residual gas pressure of $\sim5\times10^{-7}$ mbar will be improved for experimental runs in the UCNB and Nab spectrometers.

This detection system has also been used to demonstrate an important systematic consideration for experiments which take advantage of finite detectors in magnetic spectrometers, in a recent study of the non-monotonic point-spread function of monoenergetic particles in a homogeneous magnetic field~\cite{Sjue15}.

\section{Conclusion}

We have described a system designed for coincident detection of the proton and electron from neutron $\beta$ decay, using 1.5 and 2~mm thick silicon detectors with 100~nm thick dead layer, and active area diameter of 11.5~cm.  The detection system was successfully tested at LANSCE.  The individual pixels meet specifications of $\sim$3~keV FWHM energy resolution and energy threshold~$<$10~keV, with $\sim$50~ns rise time. This demonstration has been used to finalize the design of a fully instrumented detection system that will be implemented by the Nab and UCNB experiments.

\section{Appendix.  Electronic noise coefficients}

Electronics noise coefficients and their uncertainties extracted from fitting the shaping time scan data to Eq.~\ref{eq:enc} are listed in Table~\ref{tab:shapescan}.  Coefficients were extracted from $^{113}$Sn X-rays or $^{139}$Ce X-rays(*).

\begin{table}[h]
\caption{Noise coefficients per detector pixel. The pixel column follows the convention of Fig.~\ref{fig:shapetable}}  \label{tab:shapescan}
\centering
\begin{tabular}{cccc}
\hline
      pixel & h1 [$e^2 \mu s$ ] & h2  [$e^2$] & h3 $\left[\frac{e^2}{\mu s}\right]$ \\
      & ($\times10^4$)& ($\times10^4$)& ($\times10^4$)  \\
      \hline
      39W & 1.0$\pm$0.2 & 8.2$\pm$0.8 & 7.0$\pm$0.5 \\
      40W & 0.9$\pm$0.2 & 9.2$\pm$1.3 & 7.7$\pm$0.7 \\
      41W & 1.2$\pm$0.3 & 6.9$\pm$1.4 & 7.0$\pm$0.8 \\
      50W & 1.1$\pm$0.3 & 7.2$\pm$1.5 & 6.8$\pm$0.8 \\
      52W & 0.9$\pm$0.2 & 8.9$\pm$1.2 & 6.9$\pm$0.7 \\
      53W & 1.0$\pm$0.3 & 9.3$\pm$1.4 & 7.0$\pm$0.8 \\
      62W & 1.3$\pm$0.3 & 8.4$\pm$1.3 & 8.0$\pm$0.8 \\
      63W & 0.9$\pm$0.2  & 6.7$\pm$0.9  & 5.9$\pm$0.5 \\
      64W & 1.1$\pm$0.3  & 9.2$\pm$1.3  & 6.8$\pm$0.7 \\
      65W* & 2.4$\pm$0.6 & 12.4$\pm$3.0 & 7.7$\pm$1.7 \\
      66W* & 2.2$\pm$0.5 & 10.5$\pm$2.2 & 8.8$\pm$1.3 \\
      75W & 1.3$\pm$0.3 & 6.1$\pm$1.3 & 6.4$\pm$0.8 \\
      78W* & 1.8$\pm$0.5 & 10.0$\pm$2.5 & 6.1$\pm$1.4 \\
      87W* & 1.5$\pm$0.3 & 9.3$\pm$1.2 & 4.4$\pm$0.7 \\
      88W* & 2.0$\pm$0.5 & 8.2$\pm$2.3 & 7.7$\pm$1.3 \\
      89W* & 1.8$\pm$0.3 & 8.8$\pm$1.6 & 5.2$\pm$0.9 \\

      39E & 0.9$\pm$0.2 & 6.7$\pm$1.0 & 6.0$\pm$0.5  \\
      40E & 1.0$\pm$0.3  & 7.5$\pm$1.4 & 5.8$\pm$0.8 \\
      41E & 1.1$\pm$0.2  & 7.0$\pm$1.2  & 6.3$\pm$0.7 \\
      50E & 1.1$\pm$0.3  & 7.6$\pm$1.4  & 5.8$\pm$0.8 \\
      51E & 0.8$\pm$0.2  & 6.4$\pm$1.2  & 6.3$\pm$0.7 \\
      53E & 1.1$\pm$0.2  & 6.2$\pm$1.1  & 5.9$\pm$0.6 \\
      62E & 0.9$\pm$0.2  & 7.8$\pm$1.1  & 6.1$\pm$0.6 \\
      63E* & 2.3$\pm$0.4 & 8.7$\pm$2.0 & 5.5$\pm$1.2 \\
      64E & 1.0$\pm$0.2 & 5.3$\pm$1.0 & 6.5$\pm$0.6 \\
      65E & 1.2$\pm$0.3 & 7.4$\pm$1.3 & 5.7$\pm$0.7 \\
      66E & 0.9$\pm$0.3 & 8.5$\pm$1.3 & 6.5$\pm$0.7 \\
      75E* & 2.0$\pm$0.3 & 7.0$\pm$1.3 & 6.1$\pm$0.7 \\
      77E* & 1.8$\pm$0.6 & 13.0$\pm$2.8 & 4.0$\pm$1.6 \\
      78E & 4.3$\pm$0.3 & 5.9$\pm$1.6 & 10.2$\pm$1.0 \\
      87E* & 2.5$\pm$0.5 & 8.9$\pm$2.4 & 3.9$\pm$1.4 \\
      88E* & 1.7$\pm$0.4  & 8.3$\pm$1.8 & 4.0$\pm$1.0 \\
      89E* & 9.8$\pm$0.7 & 1.7$\pm$3.1 & 15.0$\pm$1.8 \\
\hline
\end{tabular}
\end{table}

\section*{Acknowledgments}
We thank our UCNA collaborators from the California Institute of Technology for their contribution of the SCS magnetic spectrometer.
This work was supported by the LDRD program of Los Alamos National Laboratory [project number 20110043DR], the National Science Foundation [contract number NSF PHY-1205833, 1307426], and the U.S. Department of Energy, Office of Science, Office of Nuclear Physics [contract numbers DE-AC52-06NA25396 under proposal 2017LANLEEDM, DE-AC05-00OR22725, DE-FG02-03ER41258, DE-FG02-97ER41042, DE-SC000\-8107, DE-SC0014622] and the Office of Workforce Development for Teachers and Scientists (WDTS) under the Science Undergraduate Laboratory Internships Program (SULI).

\section*{References}
\bibliography{Rerevised_Broussard_201607_NIMA}

\begin{thebibliography}{10}
\expandafter\ifx\csname url\endcsname\relax
  \def\url#1{\texttt{#1}}\fi
\expandafter\ifx\csname urlprefix\endcsname\relax\def\urlprefix{URL }\fi
\expandafter\ifx\csname href\endcsname\relax
  \def\href#1#2{#2} \def\path#1{#1}\fi

\bibitem{Abele20081}
H.~Abele, The neutron. {I}ts properties and basic interactions, Prog. Part.
  Nucl. Phys. 60 (2008) 1 -- 81.
\newblock \href {http://dx.doi.org/10.1016/j.ppnp.2007.05.002}
  {\path{doi:10.1016/j.ppnp.2007.05.002}}.

\bibitem{RamseyMusolf20081}
M.~Ramsey-Musolf, S.~Su, Low-energy precision tests of supersymmetry, Phys.
  Rep. 456 (2008) 1 -- 88.
\newblock \href {http://dx.doi.org/10.1016/j.physrep.2007.10.001}
  {\path{doi:10.1016/j.physrep.2007.10.001}}.

\bibitem{0954-3899-36-10-104001}
J.~S. Nico, Neutron beta decay, J. Phys. G 36 (2009) 104001.
\newblock \href {http://dx.doi.org/10.1103/PhysRevC.71.055502}
  {\path{doi:10.1103/PhysRevC.71.055502}}.

\bibitem{RevModPhys.83.1111}
D.~Dubbers, M.~G. Schmidt, The neutron and its role in cosmology and particle
  physics, Rev. Mod. Phys. 83 (2011) 1111--1171.
\newblock \href {http://dx.doi.org/10.1103/RevModPhys.83.1111}
  {\path{doi:10.1103/RevModPhys.83.1111}}.

\bibitem{ANDP:ANDP201300072}
O.~Naviliat-Cuncic, M.~González-Alonso, Prospects for precision measurements
  in nuclear $\beta$ decay in the {LHC} era, Ann. Phys. (Berl.) 525 (2013)
  600--619.
\newblock \href {http://dx.doi.org/10.1002/andp.201300072}
  {\path{doi:10.1002/andp.201300072}}.

\bibitem{PhysRevD.88.073002}
A.~N. Ivanov, M.~Pitschmann, N.~I. Troitskaya, Neutron
  ${\ensuremath{\beta}}^{\mathbf{-}}$ decay as a laboratory for testing the
  {S}tandard {M}odel, Phys. Rev. D 88 (2013) 073002.
\newblock \href {http://dx.doi.org/10.1103/PhysRevD.88.073002}
  {\path{doi:10.1103/PhysRevD.88.073002}}.

\bibitem{Cirigliano201393}
V.~Cirigliano, S.~Gardner, B.~R. Holstein, Beta decays and non-standard
  interactions in the {LHC} era, Prog. Part. Nucl. Phys. 71 (2013) 93 -- 118.
\newblock \href {http://dx.doi.org/10.1016/j.ppnp.2013.03.005}
  {\path{doi:10.1016/j.ppnp.2013.03.005}}.

\bibitem{RevModPhys.87.1483}
K.~K. Vos, H.~W. Wilschut, R.~G.~E. Timmermans, Symmetry violations in nuclear
  and neutron $\ensuremath{\beta}$ decay, Rev. Mod. Phys. 87 (2015) 1483--1516.
\newblock \href {http://dx.doi.org/10.1103/RevModPhys.87.1483}
  {\path{doi:10.1103/RevModPhys.87.1483}}.

\bibitem{Materne2009176}
S.~Materne, et~al., {PEN}e{LOPE}--on the way towards a new neutron lifetime
  experiment with magnetic storage of ultra-cold neutrons and proton
  extraction, Nucl. Instr. Meth. Phys. Res. A 611 (2009) 176 -- 180.
\newblock \href {http://dx.doi.org/10.1016/j.nima.2009.07.055}
  {\path{doi:10.1016/j.nima.2009.07.055}}.

\bibitem{PhysRevLett.111.222501}
A.~T. Yue, et~al., Improved determination of the neutron lifetime, Phys. Rev.
  Lett. 111 (2013) 222501.
\newblock \href {http://dx.doi.org/10.1103/PhysRevLett.111.222501}
  {\path{doi:10.1103/PhysRevLett.111.222501}}.

\bibitem{PhysRevC.89.052501}
D.~J. Salvat, et~al., Storage of ultracold neutrons in the
  magneto-gravitational trap of the {UCN}$\tau$ experiment, Phys. Rev. C 89
  (2014) 052501.
\newblock \href {http://dx.doi.org/10.1103/PhysRevC.89.052501}
  {\path{doi:10.1103/PhysRevC.89.052501}}.

\bibitem{Ezhov2014}
V.~F. Ezhov, et~al., Measurement of the neutron lifetime with ultra-cold
  neutrons stored in a magneto-gravitational trap (2014).
\newblock \href {http://arxiv.org/abs/1412.7434} {\path{arXiv:1412.7434}}.

\bibitem{Arimoto2015187}
Y.~Arimoto, et~al., Development of time projection chamber for precise neutron
  lifetime measurement using pulsed cold neutron beams, Nucl. Instr. Meth.
  Phys. Res. A 799 (2015) 187 -- 196.
\newblock \href {http://dx.doi.org/10.1016/j.nima.2015.08.006}
  {\path{doi:10.1016/j.nima.2015.08.006}}.

\bibitem{Arzumanov201579}
S.~Arzumanov, et~al., A measurement of the neutron lifetime using the method of
  storage of ultracold neutrons and detection of inelastically up-scattered
  neutrons, Phys. Lett. B 745 (2015) 79 -- 89.
\newblock \href {http://dx.doi.org/10.1016/j.physletb.2015.04.021}
  {\path{doi:10.1016/j.physletb.2015.04.021}}.

\bibitem{PhysRevC.94.045502}
K.~K.~H. Leung, P.~Geltenbort, S.~Ivanov, F.~Rosenau, O.~Zimmer, Neutron
  lifetime measurements and effective spectral cleaning with an ultracold
  neutron trap using a vertical {H}albach octupole permanent magnet array,
  Phys. Rev. C 94 (2016) 045502.
\newblock \href {http://dx.doi.org/10.1103/PhysRevC.94.045502}
  {\path{doi:10.1103/PhysRevC.94.045502}}.

\bibitem{1742-6596-340-1-012048}
G.~Konrad, et~al., Neutron decay with {PERC}: a progress report, J. Phys.:
  Conf. Ser. 340 (2012) 012048.
\newblock \href {http://dx.doi.org/10.1088/1742-6596/340/1/012048}
  {\path{doi:10.1088/1742-6596/340/1/012048}}.

\bibitem{PhysRevLett.110.172502}
D.~Mund, et~al., Determination of the weak axial vector coupling
  $\lambda$=${g}_{A}/{g}_{V}$ from a measurement of the $\beta$-asymmetry
  parameter {A} in neutron beta decay, Phys. Rev. Lett. 110 (2013) 172502.
\newblock \href {http://dx.doi.org/10.1103/PhysRevLett.110.172502}
  {\path{doi:10.1103/PhysRevLett.110.172502}}.

\bibitem{PhysRevC.87.032501}
M.~P. Mendenhall, et~al., Precision measurement of the neutron $\beta$-decay
  asymmetry, Phys. Rev. C 87 (2013) 032501.
\newblock \href {http://dx.doi.org/10.1103/PhysRevC.87.032501}
  {\path{doi:10.1103/PhysRevC.87.032501}}.

\bibitem{Markisch2009216}
B.~M{\"a}rkisch, et~al., The new neutron decay spectrometer {P}erkeo {III},
  Nucl. Instr. Meth. Phys. Res. A 611 (2009) 216 -- 218.
\newblock \href {http://dx.doi.org/10.1016/j.nima.2009.07.066}
  {\path{doi:10.1016/j.nima.2009.07.066}}.

\bibitem{Wietfeldt2009207}
F.~Wietfeldt, et~al., a{CORN}: An experiment to measure the
  electron–antineutrino correlation in neutron decay, Nucl. Instr. Meth.
  Phys. Res. A 611 (2009) 207 -- 210.
\newblock \href {http://dx.doi.org/10.1016/j.nima.2009.07.064}
  {\path{doi:10.1016/j.nima.2009.07.064}}.

\bibitem{Simson2009203}
M.~Simson, et~al., Measuring the proton spectrum in neutron decay--latest
  results with a{SPECT}, Nucl. Instr. Meth. Phys. Res. A 611 (2009) 203 -- 206.
\newblock \href {http://dx.doi.org/10.1016/j.nima.2009.07.068}
  {\path{doi:10.1016/j.nima.2009.07.068}}.

\bibitem{Baessler2013}
S.~Bae{\ss}ler, et~al., Neutron beta decay studies with {N}ab, AIP Conf. Proc.
  1560 (2013) 114.
\newblock \href {http://dx.doi.org/10.1063/1.4826731}
  {\path{doi:10.1063/1.4826731}}.

\bibitem{Serebrov1998}
A.~Serebrov, et~al., Measurement of the antineutrino escape asymmetry with
  respect to the spin of the decaying neutron, J. Exp. Theor. Phys. 86 (1998)
  1074--1082.
\newblock \href {http://dx.doi.org/10.1134/1.558574}
  {\path{doi:10.1134/1.558574}}.

\bibitem{Kreuz2005263}
M.~Kreuz, et~al., A measurement of the antineutrino asymmetry {B} in free
  neutron decay, Phys. Lett. B 619 (2005) 263 -- 270.
\newblock \href {http://dx.doi.org/10.1016/j.physletb.2005.05.074}
  {\path{doi:10.1016/j.physletb.2005.05.074}}.

\bibitem{PhysRevLett.99.191803}
M.~Schumann, et~al., Measurement of the neutrino asymmetry parameter {B} in
  neutron decay, Phys. Rev. Lett. 99 (2007) 191803.
\newblock \href {http://dx.doi.org/10.1103/PhysRevLett.99.191803}
  {\path{doi:10.1103/PhysRevLett.99.191803}}.

\bibitem{Broussard2012}
L.~J. Broussard, et~al., {UCNB}: The neutrino asymmetry in polarized ultracold
  neutron decay, AIP Conf. Proc. 1560 (2013) 149.
\newblock \href {http://dx.doi.org/10.1063/1.4826741}
  {\path{doi:10.1063/1.4826741}}.

\bibitem{PhysRevD.85.054512}
T.~Bhattacharya, et~al., Probing novel scalar and tensor interactions from
  (ultra)cold neutrons to the {LHC}, Phys. Rev. D 85 (2012) 054512.
\newblock \href {http://dx.doi.org/10.1103/PhysRevD.85.054512}
  {\path{doi:10.1103/PhysRevD.85.054512}}.

\bibitem{PhysRevC.87.065504}
S.~Gardner, B.~Plaster, Framework for maximum likelihood analysis of neutron
  $\ensuremath{\beta}$ decay observables to resolve the limits of the
  ${V}\ensuremath{-}{A}$ law, Phys. Rev. C 87 (2013) 065504.
\newblock \href {http://dx.doi.org/10.1103/PhysRevC.87.065504}
  {\path{doi:10.1103/PhysRevC.87.065504}}.

\bibitem{PhysRevC.88.048501}
R.~W. Pattie, K.~P. Hickerson, A.~R. Young, Limits on tensor coupling from
  neutron $\ensuremath{\beta}$ decay, Phys. Rev. C 88 (2013) 048501.
\newblock \href {http://dx.doi.org/10.1103/PhysRevC.88.048501}
  {\path{doi:10.1103/PhysRevC.88.048501}}.

\bibitem{PhysRevC.92.069902}
R.~W. Pattie, K.~P. Hickerson, A.~R. Young, Erratum: {L}imits on tensor
  coupling from neutron $\ensuremath{\beta}$ decay [{P}hys. {R}ev. {C}
  \textbf{88} , 048501 (2013)], Phys. Rev. C 92 (2015) 069902.
\newblock \href {http://dx.doi.org/10.1103/PhysRevC.92.069902}
  {\path{doi:10.1103/PhysRevC.92.069902}}.

\bibitem{PhysRevC.89.025501}
F.~Wauters, A.~Garc\'{\i}a, R.~Hong, Limits on tensor-type weak currents from
  nuclear and neutron $\ensuremath{\beta}$ decays, Phys. Rev. C 89 (2014)
  025501.
\newblock \href {http://dx.doi.org/10.1103/PhysRevC.89.025501}
  {\path{doi:10.1103/PhysRevC.89.025501}}.

\bibitem{PhysRevC.91.049904}
F.~Wauters, A.~Garc\'{\i}a, R.~Hong, Erratum: {L}imits on tensor-type weak
  currents from nuclear and neutron $\ensuremath{\beta}$ decays [{P}hys. {R}ev.
  {C} \textbf{89} , 025501 (2014)], Phys. Rev. C 91 (2015) 049904.
\newblock \href {http://dx.doi.org/10.1103/PhysRevC.91.049904}
  {\path{doi:10.1103/PhysRevC.91.049904}}.

\bibitem{Yerozolimsky2000491}
B.~G. Yerozolimsky, Free neutron decay: a review of the contemporary situation,
  Nucl. Instr. Meth. Phys. Res. A 440~(3) (2000) 491 -- 498.
\newblock \href {http://dx.doi.org/10.1016/S0168-9002(99)01026-8}
  {\path{doi:10.1016/S0168-9002(99)01026-8}}.

\bibitem{Abele2000499}
H.~Abele, The standard model and the neutron $\beta$-decay, Nucl. Instr. Meth.
  Phys. Res. A 440~(3) (2000) 499 -- 510.
\newblock \href {http://dx.doi.org/10.1016/S0168-9002(99)01027-X}
  {\path{doi:10.1016/S0168-9002(99)01027-X}}.

\bibitem{spivak}
P.~E. Spivak,
  \href{http://www.jetp.ac.ru/cgi-bin/e/index/e/67/9/p1735?a=list}{Neutron
  lifetime obtained from {A}tomic-{E}nergy-{I}nstitute experiment [{J}. {E}xp.
  {T}heor. {P}hys. {L}ett. 28, 303 (1978)]}, J. Exp. Theor. Phys. 67 (1988)
  1735, [Z. Eksp. Teor. Fiz. {\bf 94}, 1 (1988)].
\newline\urlprefix\url{http://www.jetp.ac.ru/cgi-bin/e/index/e/67/9/p1735?a=list}

\bibitem{EROZOLIMSKII199133}
B.~Erozolimskii, I.~Kuznetsov, I.~Stepanenko, I.~Kuida, Y.~Mostovoi, New
  measurements of the electron-neutron spin asymmetry in neutron beta-decay,
  Phys. Lett. B 263 (1991) 33 -- 38.
\newblock \href {http://dx.doi.org/10.1016/0370-2693(91)91703-X}
  {\path{doi:10.1016/0370-2693(91)91703-X}}.

\bibitem{PhysRevLett.33.41}
R.~I. Steinberg, P.~Liaud, B.~Vignon, V.~W. Hughes, New experimental limit on
  {T} invariance in polarized-neutron $\ensuremath{\beta}$ decay, Phys. Rev.
  Lett. 33 (1974) 41--44.
\newblock \href {http://dx.doi.org/10.1103/PhysRevLett.33.41}
  {\path{doi:10.1103/PhysRevLett.33.41}}.

\bibitem{Mostovoi2001}
Y.~A. Mostovoi, et~al., Experimental value of {GA}/{GV} from a measurement of
  both {P}-odd correlations in free-neutron decay, Phys. Atom. Nucl. 64 (2001)
  1955--1960.
\newblock \href {http://dx.doi.org/10.1134/1.1423745}
  {\path{doi:10.1134/1.1423745}}.

\bibitem{Beck2002}
M.~Beck, et~al., The upper limit of the branching ratio for radiative beta
  decay of free neutrons, J. Exp. Theor. Phys. Lett. 76 (2002) 332--336,
  [Pis'ma Z. Eksp. Teor. Fiz. {\bf 76}, 392 (2002)].
\newblock \href {http://dx.doi.org/10.1134/1.1525031}
  {\path{doi:10.1134/1.1525031}}.

\bibitem{Soldner200449}
T.~Soldner, L.~Beck, C.~Plonka, K.~Schreckenbach, O.~Zimmer, New limit on {T}
  violation in neutron decay, Phys. Lett. B 581 (2004) 49 -- 55.
\newblock \href {http://dx.doi.org/10.1016/j.physletb.2003.12.004}
  {\path{doi:10.1016/j.physletb.2003.12.004}}.

\bibitem{PhysRevD.18.3970}
C.~Stratowa, R.~Dobrozemsky, P.~Weinzierl, Ratio $|\frac{{g}_{A}}{{g}_{V}}|$
  derived from the proton spectrum in free-neutron decay, Phys. Rev. D 18
  (1978) 3970--3979.
\newblock \href {http://dx.doi.org/10.1103/PhysRevD.18.3970}
  {\path{doi:10.1103/PhysRevD.18.3970}}.

\bibitem{PhysRevLett.100.151801}
M.~Schumann, et~al., Measurement of the proton asymmetry parameter in neutron
  beta decay, Phys. Rev. Lett. 100 (2008) 151801.
\newblock \href {http://dx.doi.org/10.1103/PhysRevLett.100.151801}
  {\path{doi:10.1103/PhysRevLett.100.151801}}.

\bibitem{Hoedl2006}
S.~A. Hoedl, A.~R. Young, H.~Ade, A.~Lozano, An electron transparent proton
  detector for neutron decay studies, J. Appl. Phys. 99 (2006) 084904.
\newblock \href {http://dx.doi.org/10.1063/1.2186970}
  {\path{doi:10.1063/1.2186970}}.

\bibitem{1748-0221-11-02-C02068}
D.~Gaisbauer, Y.~Bai, I.~Konorov, S.~Paul, D.~Steffen, Self-triggering readout
  system for the neutron lifetime experiment {PEN}e{LOPE}, J. Instrum. 11
  (2016) C02068.
\newblock \href {http://dx.doi.org/10.1088/1748-0221/11/02/C02068}
  {\path{doi:10.1088/1748-0221/11/02/C02068}}.

\bibitem{0295-5075-33-3-187}
J.~Byrne, et~al., A revised value for the neutron lifetime measured using a
  {P}enning trap, Europhys. Lett. 33 (1996) 187.
\newblock \href {http://dx.doi.org/10.1209/epl/i1996-00319-x}
  {\path{doi:10.1209/epl/i1996-00319-x}}.

\bibitem{0954-3899-28-6-314}
J.~Byrne, et~al., Determination of the electron--antineutrino angular
  correlation coefficient $a_0$ and the parameter $|\lambda| = |{G}_{A} /
  {G}_{V}|$ in free neutron $\beta$-decay from measurements of the integrated
  energy spectrum of recoil protons stored in an ion trap, J. Phys. G 28 (2002)
  1325.
\newblock \href {http://dx.doi.org/10.1088/0954-3899/28/6/314}
  {\path{doi:10.1088/0954-3899/28/6/314}}.

\bibitem{PhysRevC.71.055502}
J.~S. Nico, et~al., Measurement of the neutron lifetime by counting trapped
  protons in a cold neutron beam, Phys. Rev. C 71 (2005) 055502.
\newblock \href {http://dx.doi.org/10.1103/PhysRevC.71.055502}
  {\path{doi:10.1103/PhysRevC.71.055502}}.

\bibitem{PhysRevC.81.035503}
R.~L. Cooper, et~al., Radiative $\ensuremath{\beta}$ decay of the free neutron,
  Phys. Rev. C 81 (2010) 035503.
\newblock \href {http://dx.doi.org/10.1103/PhysRevC.81.035503}
  {\path{doi:10.1103/PhysRevC.81.035503}}.

\bibitem{PhysRevC.86.035505}
T.~E. Chupp, et~al., Search for a {T}-odd, {P}-even triple correlation in
  neutron decay, Phys. Rev. C 86 (2012) 035505.
\newblock \href {http://dx.doi.org/10.1103/PhysRevC.86.035505}
  {\path{doi:10.1103/PhysRevC.86.035505}}.

\bibitem{Amsbaugh201540}
J.~Amsbaugh, et~al., Focal-plane detector system for the {KATRIN} experiment,
  Nucl. Instr. Meth. Phys. Res. A 778 (2015) 40 -- 60.
\newblock \href {http://dx.doi.org/10.1016/j.nima.2014.12.116}
  {\path{doi:10.1016/j.nima.2014.12.116}}.

\bibitem{Wilburn2009}
W.~S. Wilburn, et~al.,
  \href{http://rmf.smf.mx/pdf/rmf-s/55/2/55_2_119.pdf}{Measurement of the
  neutrino-spin correlation parameter {B} in neutron decay using ultracold
  neutrons}, Rev. Mex. F{\'i}s. S 55 (2009) 119.
\newline\urlprefix\url{http://rmf.smf.mx/pdf/rmf-s/55/2/55_2_119.pdf}

\bibitem{0954-3899-41-11-114007}
A.~R. Young, et~al., Beta decay measurements with ultracold neutrons: a review
  of recent measurements and the research program at {L}os {A}lamos {N}ational
  {L}aboratory, J. Phys. G 41 (2014) 114007.
\newblock \href {http://dx.doi.org/10.1088/0954-3899/41/11/114007}
  {\path{doi:10.1088/0954-3899/41/11/114007}}.

\bibitem{Bowman2005}
J.~D. Bowman, On the measurement of the electron-neutrino correlation in
  neutron beta decay, J. Res. Natl. Inst. Stand. Tech. 110 (2005) 407.
\newblock \href {http://dx.doi.org/10.6028/jres.110.061}
  {\path{doi:10.6028/jres.110.061}}.

\bibitem{Pocanic2009211}
D.~Po{\v c}ani{\' c}, et~al., Nab: Measurement principles, apparatus and
  uncertainties, Nucl. Instr. Meth. Phys. Res. A 611 (2009) 211 -- 215.
\newblock \href {http://dx.doi.org/10.1016/j.nima.2009.07.065}
  {\path{doi:10.1016/j.nima.2009.07.065}}.

\bibitem{0954-3899-41-11-114003}
S.~Bae{\ss}ler, J.~D. Bowman, S.~Penttil{\"a}, D.~Po{\v c}ani{\' c}, New
  precision measurements of free neutron beta decay with cold neutrons, J.
  Phys. G 41~(11) (2014) 114003.
\newblock \href {http://dx.doi.org/10.1088/0954-3899/41/11/114003}
  {\path{doi:10.1088/0954-3899/41/11/114003}}.

\bibitem{Saunders2013}
A.~Saunders, et~al., Performance of the {L}os {A}lamos {N}ational {L}aboratory
  spallation-driven solid-deuterium ultra-cold neutron source, Rev. Sci.
  Instrum. 84 (2013) 013304.
\newblock \href {http://dx.doi.org/10.1063/1.4770063}
  {\path{doi:10.1063/1.4770063}}.

\bibitem{PhysRevC.86.055501}
B.~Plaster, et~al., Measurement of the neutron $\beta$-asymmetry parameter
  ${A}_{0}$ with ultracold neutrons, Phys. Rev. C 86 (2012) 055501.
\newblock \href {http://dx.doi.org/10.1103/PhysRevC.86.055501}
  {\path{doi:10.1103/PhysRevC.86.055501}}.

\bibitem{Micron}
{Micron Semiconductor Limited, UK}, \url{http://micronsemiconductor.co.uk}.

\bibitem{SalasBacci2014408}
A.~Salas-Bacci, et~al., Characterization of large area, thick, and segmented
  silicon detectors for neutron experiments, Nucl. Instr. Meth. Phys. Res. A
  735 (2014) 408 -- 415.
\newblock \href {http://dx.doi.org/10.1016/j.nima.2013.09.059}
  {\path{doi:10.1016/j.nima.2013.09.059}}.

\bibitem{NI}
{National Instruments PXIe-5171R},
  \url{http://sine.ni.com/nips/cds/view/p/lang/en/nid/212657}.

\bibitem{spieler2005}
H.~Spieler, Semiconductor Detector Systems, Oxford University Press, Oxford,
  England, 2005.

\bibitem{knoll}
G.~F. Knoll, Radiation Detection and Measurement, Wiley, Hoboken, NJ, 2000.

\bibitem{PEHL196845}
R.~Pehl, F.~Goulding, D.~Landis, M.~Lenzlinger, Accurate determination of the
  ionization energy in semiconductor detectors, Nucl. Instr. Meth. 59 (1968) 45
  -- 55.
\newblock \href {http://dx.doi.org/10.1016/0029-554X(68)90342-X}
  {\path{doi:10.1016/0029-554X(68)90342-X}}.

\bibitem{JORDANOV1994261}
V.~T. Jordanov, G.~F. Knoll, A.~C. Huber, J.~A. Pantazis, Digital techniques
  for real-time pulse shaping in radiation measurements, Nucl. Instr. Meth.
  Phys. Res. A 353 (1994) 261 -- 264.
\newblock \href {http://dx.doi.org/10.1016/0168-9002(94)91652-7}
  {\path{doi:10.1016/0168-9002(94)91652-7}}.

\bibitem{Bertuccio93}
G.~Bertuccio, A.~Pullia, A method for the determination of the noise parameters
  in preamplifying systems for semiconductor radiation detectors, Rev. Sci.
  Instrum. 64 (1993) 3294--3298.
\newblock \href {http://dx.doi.org/10.1063/1.1144293}
  {\path{doi:10.1063/1.1144293}}.

\bibitem{salvat2006penelope}
F.~Salvat, J.~M. Fern{\'a}ndez-Varea, J.~Sempau,
  \href{http://www.oecd-nea.org/globalsearch/download.php?doc=77434}{{PENELOPE}
  2011: A code system for {M}onte {C}arlo simulation of electron and photon
  transport}, in: OECD Nuclear Energy Agency, Issy-les-Moulineaux, France,
  2011.
\newline\urlprefix\url{http://www.oecd-nea.org/globalsearch/download.php?doc=77434}

\bibitem{NNDC}
Data retrieved from the {N}ational {N}uclear {D}ata {C}enter {W}orld{W}ide{W}eb
  site, \url{http://www.nndc.bnl.gov/}, accessed: 2016-06-01.

\bibitem{Holley12}
A.~T. Holley, et~al., A high-field adiabatic fast passage ultracold neutron
  spin flipper for the {UCNA} experiment, Rev. Sci. Instrum. 83 (2012) 073505.
\newblock \href {http://dx.doi.org/10.1063/1.4732822}
  {\path{doi:10.1063/1.4732822}}.

\bibitem{Morris2009248}
C.~Morris, et~al., Multi-wire proportional chamber for ultra-cold neutron
  detection, Nucl. Instr. Meth. Phys. Res. A 599 (2009) 248 -- 250.
\newblock \href {http://dx.doi.org/10.1016/j.nima.2008.11.099}
  {\path{doi:10.1016/j.nima.2008.11.099}}.

\bibitem{Wang201530}
Z.~Wang, et~al., A multilayer surface detector for ultracold neutrons, Nucl.
  Instr. Meth. Phys. Res. A 798 (2015) 30 -- 35.
\newblock \href {http://dx.doi.org/10.1016/j.nima.2015.07.010}
  {\path{doi:10.1016/j.nima.2015.07.010}}.

\bibitem{Agostinelli2003250}
S.~Agostinelli, et~al., Geant4--a simulation toolkit, Nucl. Instr. Meth. Phys.
  Res. A 506~(3) (2003) 250 -- 303.
\newblock \href {http://dx.doi.org/10.1016/S0168-9002(03)01368-8}
  {\path{doi:10.1016/S0168-9002(03)01368-8}}.

\bibitem{ZiegetalSRIM2010}
J.~F. Ziegler, J.~P. Biersack, M.~D. Ziegler, \href{www.srim.org}{The
  {S}topping and {R}ange of {I}ons in {M}atter}, {SRIM}-2013 (2013).
\newline\urlprefix\url{www.srim.org}

\bibitem{Sjue15}
S.~K.~L. Sjue, et~al., Radial distribution of charged particles in a magnetic
  field, Rev. Sci. Instrum. 86 (2015) 023102.
\newblock \href {http://dx.doi.org/10.1063/1.4906547}
  {\path{doi:10.1063/1.4906547}}.

\end{thebibliography}
\biboptions{sort&compress}

\end{document}